\documentclass{PoS}
\usepackage{lineno}
\usepackage{float}
\title{Searches for gluinos and squarks}

\ShortTitle{Searches for gluinos and squarks}

\author{\speaker{Tamas Almos Vami} for the ATLAS and CMS Collaborations\\
        Wigner Research Centre for Physics\\
        E-mail: \email{tamas.almos.vami@cern.ch}}

\abstract{Despite the great success of the Standard Model, it still does not explain Dark Matter, matter/antimatter asymmetry and it does not unite the electroweak and the strong forces. As a possible solution, the theory of Supersymmetry was proposed. In this paper, three searches for strong production Supersymmetry are presented. The data correspond to a total integrated luminosity of 137 fb$^{-1}$ at a center-of-mass energy of 13 TeV, recorded at the Large Hadron Collider during the years 2015-2018. There has been no significant excess observed relative to the Standard Model predictions. The data gave the following best 95\% exclusion limits for the masses of squarks and gluinos: 2250 GeV, 1260 GeV and 1225 GeV which are obtained from the inclusive search for gluinos, bottom squarks and top squarks, respectively. The search for disappearing tracks extends the gluino mass limit to as much as 2460 GeV, and the neutralino mass limit to as much as 2000 GeV.}

\FullConference{7th Conference of Large Hadron Collider Physics\\
		(LHCP2019)\\
		20 - 25 May, 2019\\
		Puebla, Mexico}

\begin{document}

\section{Introduction}
The Standard Model of Particle Physics (SM) cannot be the final theory of the Universe since it does not explain the matter/antimatter asymmetry, it does not provide any candidate for Dark Matter, it does not give any reason why the mass of the Higgs boson is so low with respect to the Planck mass and it does not unify the gauge couplings. Extending the SM with new particles (superpartners) based on a new symmetry called supersymmetry (SUSY) \cite{ZuminoSUSY} seems to solve these problems theoretically \cite{GUT,WittenSUSYBreaking,Predictions,EllisBigBang,SUSYDM}. Superpartners of the SM fermions are bosons while the SM bosons have fermionic partners. The partners of the gluons are the gluinos ($\widetilde{g}$), whereas the partners of the Higgs boson (called higgsinos) and the electro-weak gauge bosons (called winos and bino) mix to form charged and neutral mass eigenstates called charginos ($\widetilde{\chi}^\pm_{1,2}$) and neutralinos ($\widetilde{\chi}^0_{1,2,3,4}$).

In this paper, three searches for strong production SUSY \cite{ATLAS-CONF-2019-015,CMS-SUS-19-008,CMS-SUS-19-005} with data from the ATLAS and CMS detectors recorded during the years 2015-2018 are presented for an integrated luminosity of 137\,$\mathrm{fb}^{-1}$ (139\,$\mathrm{fb}^{-1}$ for \cite{ATLAS-CONF-2019-015}) with the center-of-mass energy of 13\,TeV. Based on \cite{ATLAS-CONF-2019-015} and \cite{CMS-SUS-19-008} we report on a search with two same-sign leptons or at least three leptons and jets, while based on \cite{CMS-SUS-19-005} an inclusive search and also search for long lived charginos are shown.

\section{The ATLAS and CMS detectors}
Both the ATLAS and CMS detectors are  multi-purpose particle detectors aiming to cover a wide range of high energy physics topics.

\begin{figure}[h!]
    \centering
    \includegraphics[width=\textwidth]{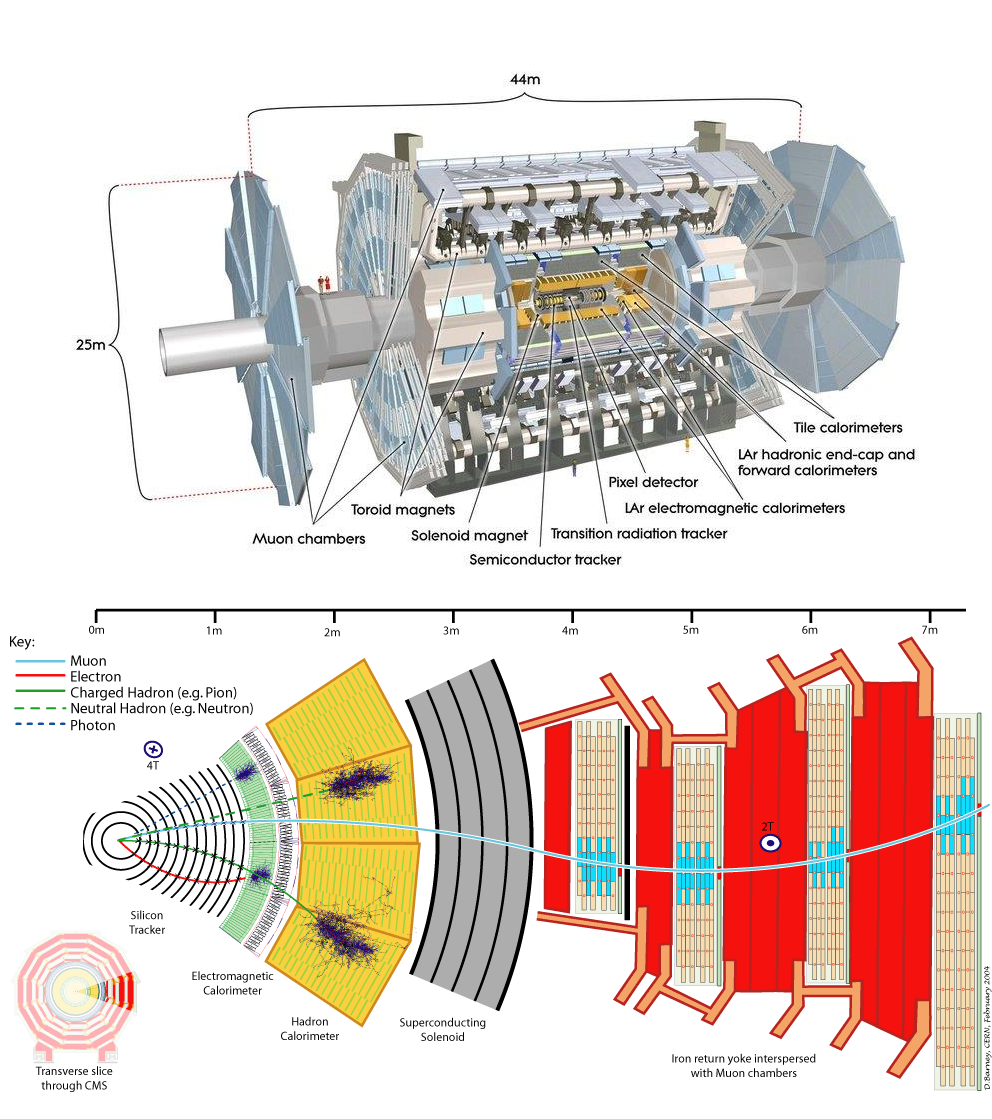}
    \caption{Schematic view of the ATLAS (top) and CMS (bottom) detectors.}
    \label{fig:ATLASandCMS}
\end{figure}

ATLAS has an inner tracking
detector surrounded by a thin superconducting solenoid providing a 2\,T axial magnetic field, a lead/liquid-argon based electromagnetic calorimeter and a steel/scintillator-tile hadron calorimeter, and a muon spectrometer (MS). The MS consists of three air-core toroidal superconducting magnets with eight coils each. The toroids' field integral ranges between 2.0 and 6.0 T$\cdot$m across the detector. Further details about the ATLAS detector can be found in \cite{ATLAS}. A schematic view of the ATLAS detector is shown on the top of Figure \ref{fig:ATLASandCMS}.

The central feature of the CMS apparatus is a superconducting solenoid of 6\,m internal diameter, providing a magnetic field of 3.8\,T. Within the solenoid volume are a silicon pixel and strip tracker, a lead tungstate crystal electromagnetic calorimeter, and a brass and scintillator hadron calorimeter, each composed of a barrel and two endcap sections. Forward calorimeters extend the pseudorapidity coverage provided by the barrel and endcap detectors. Muons are detected in gas-ionization chambers embedded in the steel flux-return yoke outside the solenoid. A more detailed description of the CMS detector, together with a definition of the coordinate system used and the relevant kinematic variables, can be found in \cite{Chatrchyan:2008zzk}. A schematic view of the CMS detector is shown on the bottom of Figure \ref{fig:ATLASandCMS}.

In case of both detectors, events of interest are selected using a two-tiered trigger system. The first level, composed of custom hardware processors, uses information from the calorimeters and muon detectors to select events at a rate of around 100 {kHz} within a time interval of less than 4\,$\mu$s. The second level, known as the high-level trigger, consists of a farm of processors running a version of the full event reconstruction software optimized for fast processing, and reduces the event rate to around 1\,kHz before data storage.~\cite{Khachatryan:2016bia}

\newpage
\section{Object reconstruction}


ATLAS uses the \textsc{FastJet} library \cite{FastJet} to reconstruct jets with $p_T$ higher than 20 GeV up to |$\eta$| = 4.9, relying on topological energy clusters in the calorimeter at the EM scale. Jets originating from b quarks (b-jets) are identified in the |$\eta$| < 2.5 region  by the MV2c10 b-tagging algorithm \cite{btaggingATLAS}.

CMS uses the particle-flow (PF) algorithm~\cite{CMS-PRF-14-001} which reconstructs and identifies each individual particle in an event, with an optimized combination of information from the various elements of the CMS detector. Jets are clustered from PF candidates using the anti-kT algorithm with a distance parameter of 0.4. In \cite{CMS-SUS-19-005} only jets with $p_T$ higher than 30 GeV were considered while in \cite{CMS-SUS-19-008} the threshold was set to 40 GeV. b-jets are
identified using a deep neural network algorithm \cite{btagdeep}.

The missing transverse momentum vector $\vec{p}_T^{miss}$ is defined as the projection of the negative vector sum of the momenta of all reconstructed PF candidates on the plane perpendicular to the beams. Its magnitude is referred to as MET.

$H_T$ is defined as the sum of all jets' absolute values of momenta, while missing $H_T$, or MHT, is the absolute value of the negative vector sum for the jets. Effective mass is then understood as the sum of the $H_T$ and MHT.

As a discriminator, the $M_{T2}$ is used with the definition of
\begin{equation}
    M_{T2} = \min_{MET(X1) +MET(X2)=MET}[\max(M_T^{(1)}),(M_T^{(2)})]
\end{equation}
where $M_T^{(i)}$ are the transverse masses of the pseudo-jets. Pseudo-jets are obtained by jet pairing with the largest dijet invariant mass and iteratively clustering all selected jets until two stable jets are left. The minimization is performed over all trial momenta satisfying the MET constraint, where MET(X1) and MET(X2) are the vectors decomposed to the pseudo-jets (denoted with X1 and X2 in this case).

\section{Event categorization, backgrounds and systematic uncertainties}
\subsection{Searches with same charged lepton pairs, or at least three leptons with jets and MET}
In \cite{ATLAS-CONF-2019-015} and \cite{CMS-SUS-19-008} searches, the signature is either two isolated electrons/muons with the same electric charge, or at least three isolated electrons/muons with jets and MET. The inclusive NLO cross-section of the process is in the order of 1\,pb thus same-sign leptons are suppressed by more than three orders of magnitude with respect to the production of opposite-sign lepton pairs.
\begin{itemize}
    \item The number of leptons and their relative electric charges,
    \item  the number of jets,
    \item  the number of b-tagged jets,
    \item the effective mass and
    \item  the invariant mass of same-sign electron
    \end{itemize}
are used to categorize events in the ATLAS search while in the CMS search the following extra variables
\begin{itemize}
    \item MET,
    \item HT,
    \item minimum $M_T$,
    \item and lepton $p_T$
\end{itemize}
are used, too. CMS further categorizes by the $p_T$ of the leptons: HH (high high), HL (high low), LL (low low). For the high category the $p_T$ of the lepton has to be bigger than 25 GeV.

The main background where two or more prompt leptons are present, including a genuine same-signed (SS) pairs, is coming from processes like WZ, SS W pair, or $t\bar{t}$ + X events. This is estimated using simulated samples, with correction factors applied to take into account the small data/MC differences. Processes like $t\bar{t}$+jets and W+jets contribute to the "at least one nonprompt or fake lepton" category. This is estimated from data. Events with a pair of opposite sign leptons, one of them mistakenly reconstructed with the wrong charge, so called "charge-flipped" electrons contribute about the level of 10\%. These background events were estimated based on simulations.

\begin{figure}[h!]
    \centering
    \includegraphics[width=\textwidth]{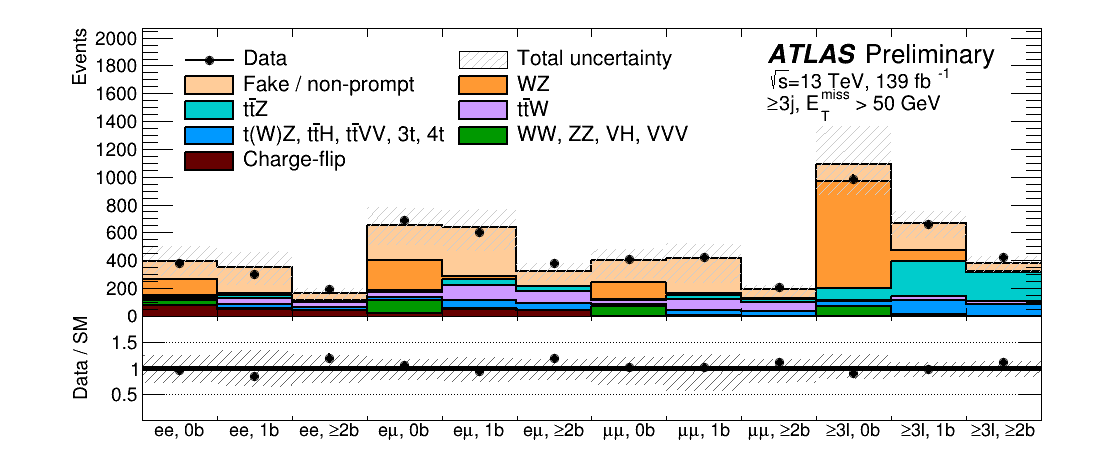}
    \caption{The x axis shows the different SRs, the box below is the ratio of events from data and MC simulation, the hashed bands are the total uncertainties \cite{ATLAS-CONF-2019-015}.}
    \label{fig:SR2}
\end{figure}

\begin{figure}[h!]
    \centering
    \includegraphics[width=.85\textwidth]{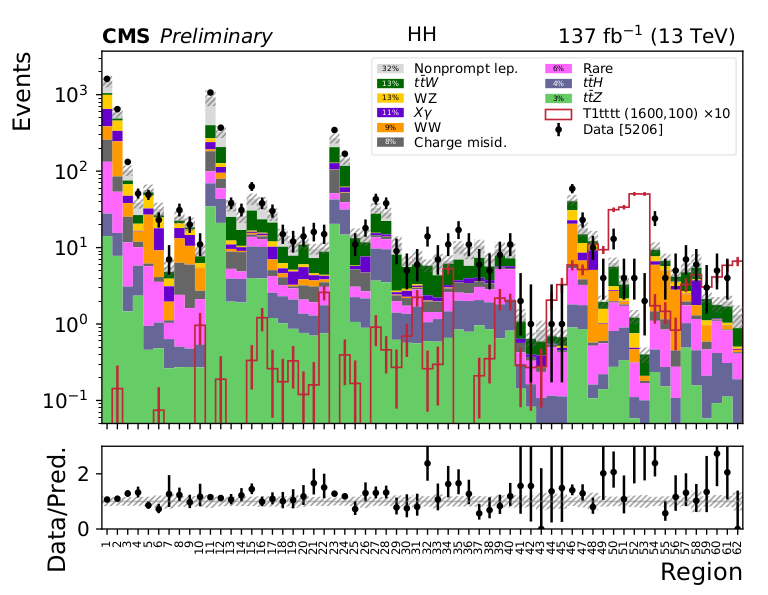}
    \caption{The x axis shows the different SRs, the box below is the ratio of events from data and MC simulation, the hashed bands are the total uncertainties. With a red curve the T1tttt yields are plotted \cite{CMS-SUS-19-008}.}
    \label{fig:SR1}
    \centering
    \includegraphics[width=\textwidth]{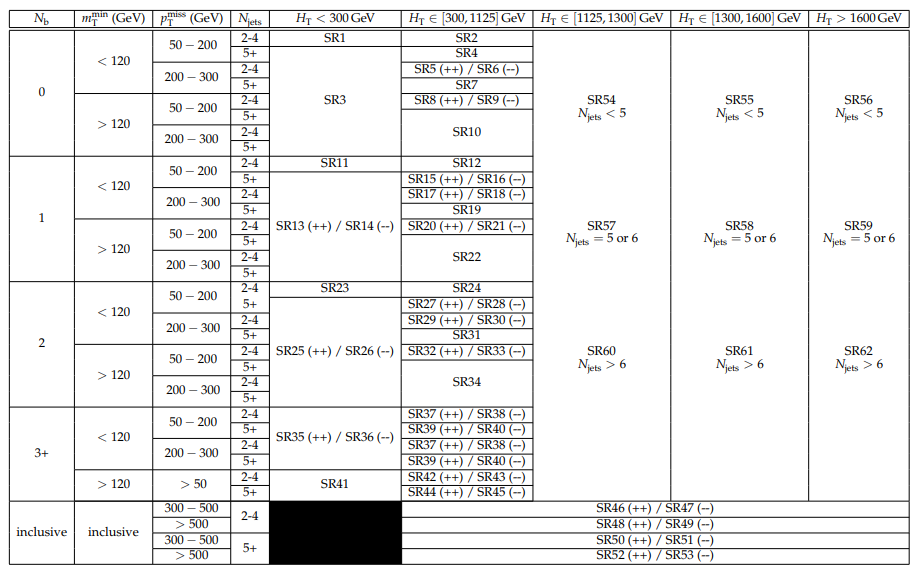}
    \caption{Definitions for the SRs in the Figure \ref{fig:SR1} \cite{CMS-SUS-19-008}.}
    \label{fig:SRDefs}
\end{figure}

In Figure \ref{fig:SR2} the x axis shows the different signal regions (SRs), the box below is the ratio of events from data and MC simulation, the hashed bands are the total uncertainties. For the CMS plot in Figure \ref{fig:SR1}, the signal T1tttt, when a gluino decays to two tops and a neutralino, is shown with a red curve.  There are 62 regions in total, the definition of each regions is shown in Figure \ref{fig:SRDefs}.

\newpage
\subsection{Inclusive $M_{T2}$ search and the search for disappearing tracks}
In \cite{CMS-SUS-19-005}, both the inclusive $M_{T2}$ search and the search for disappearing tracks, collision events are categorized by 
\begin{itemize}
    \item  the number of jets,
    \item  the number of b-tagged jets, and
    \item $H_T$.
\end{itemize}

For the inclusive search the $M_{T2}$ variable is used to further categorize, while  for the disappearing track search a short (disappearing) track is required.

The backgrounds in jets-plus-MET final states typically arise from three categories of SM
processes:
\begin{itemize}
    \item lost lepton (mostly from W+jets and $t\bar{t}$+jets events)
    \item irreducible, i.e., Z+jets events
    \item instrumental background, i.e., mostly multijet events with no genuine MET
\end{itemize}

\begin{figure}[h!]
    \centering
    \includegraphics[width=\textwidth]{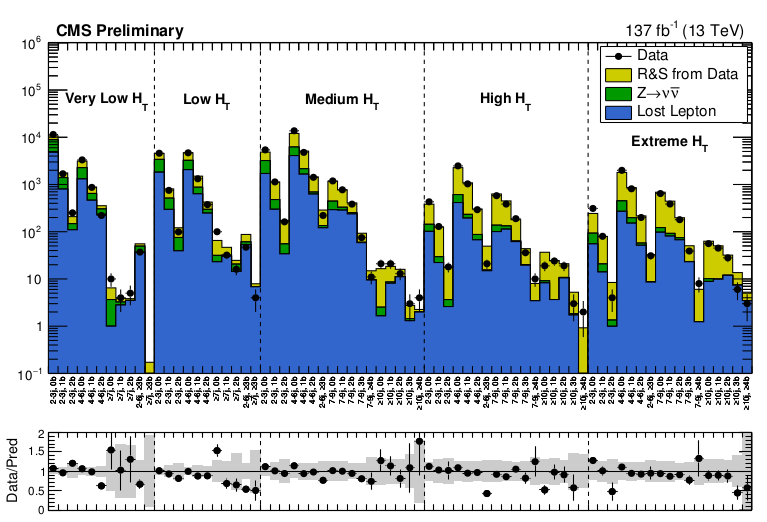}
    \caption{The x axis shows the different control regions (CRs) used to validate the R\&S method. The box below is the ratio of events from data and MC simulation, the hashed bands are the total uncertainties of the predictions \cite{CMS-SUS-19-005}.}
    \label{fig:SR3}
\end{figure}

Background is estimated from data and the rebalance and smear (R\&S) method is used to estimate multijet backgrounds. The R\&S consists of two steps. In the first step, multijet events are rebalanced by adjusting the transverse momenta of the jets such that the resulting MET is approximately null. In the smearing step, the $p_T$ of the jets in each rebalanced seed event are smeared
according to the jet response function, in order to emulate the instrumental effects. The validation of the R\&S multijet background
prediction is shown in Figure \ref{fig:SR3}, in control regions in data with
$\Delta\phi_{min}$ < 0.3. Bins on the x-axis are the
($H_T$, \#jet, \#b-jet)
topological regions. The gray band represents the total uncertainty on the prediction.

Background for the disappearing tracks search could originate from three sources when events contain disappearing tracks:
\begin{itemize}
    \item fake tracks,
    \item charged pions undergoing a significant interaction in the tracking detector or poorly reconstructed,
    \item leptons undergoing a significant interaction in the tracking detector or poorly reconstructed.
\end{itemize}

\begin{figure}[h!]
    \centering
    \includegraphics[width=.85\textwidth]{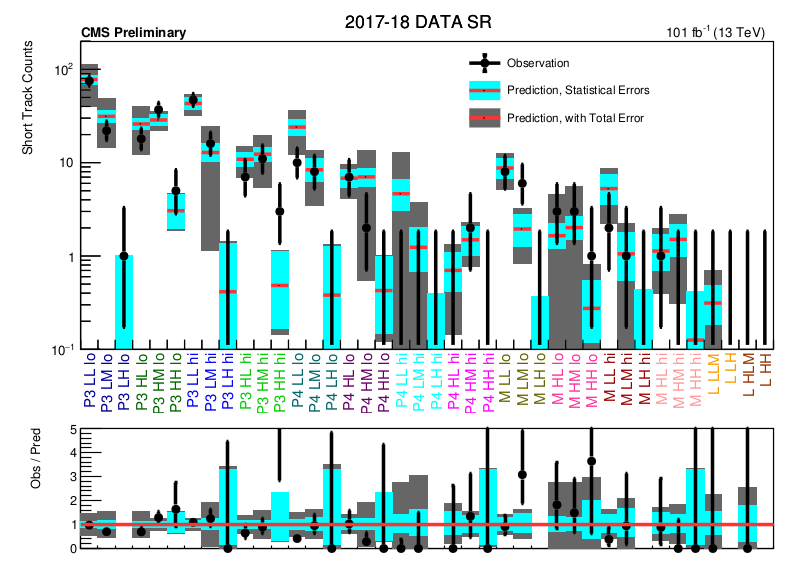}
    \caption{The x axis shows the different SRs, the box below is the ratio of events from data and MC simulation, the hashed bands are the total uncertainties \cite{CMS-SUS-19-008}.}
    \label{fig:SR4}
\end{figure}

The background is estimated from data:
$$ N_{ST}^{est} = \left(\frac{N_{ST}^{obs}}{N_{STC}^{obs}}\right)_{CR}N_{STC}^{obs} $$
where ST is for short tracks (i.e. disappearing one), STC is for short track candidates. The ratio of $N_{ST}^{obs}/N_{STC}^{obs}$ is measured in a control region independently from the region under study. The background prediction is validated in data in an intermediate $M_{T2}$ validation region (100 < $M_{T2}$ < 200 GeV), orthogonal to the high $M_{T2}$ signal region. Comparison of estimated
background and observed
data events in 2017-2018
data is shown in Figure \ref{fig:SR4}. Please note that the pixel detector was replaced in in 2017, this is why not the full dataset is shown.  A similar plot exists for 2016 as well. The black points are the actual observed data counts, the cyan band represents the statistical uncertainty on the prediction and the gray band represents the total uncertainty. Regions whose predictions use the same measurement of ${N_{ST}^{obs}}/{N_{STC}^{obs}}$ are identified by the colors of the labels. Bins with no entry in the ratio have zero predicted background. The first letter (P/M/L) defines the track length, P being pixel only track, M being a track with less than seven layers and L is for the long tracks with more than seven measurement layer. The HH/HL/LL was explained above and the hi/lo categorizes if the $p_T$ of the track is higher or lower than 50 GeV.

\newpage
\section{Results and interpretation}
The observed data yields in the signal regions are statistically consistent with the background prediction from SM processes in all cases.

Following the LHC CLs procedure \cite{CLs} upper limits were set on the production cross sections of various SUSY simplified models. 95\% confidence level exclusion limits on the production of pairs of gluinos and squarks is shown in the following figures as a function of gluino/squark and LSP (lightest SUSY particle) masses.

\begin{figure}[h!]
    \centering
    \includegraphics[width=\textwidth]{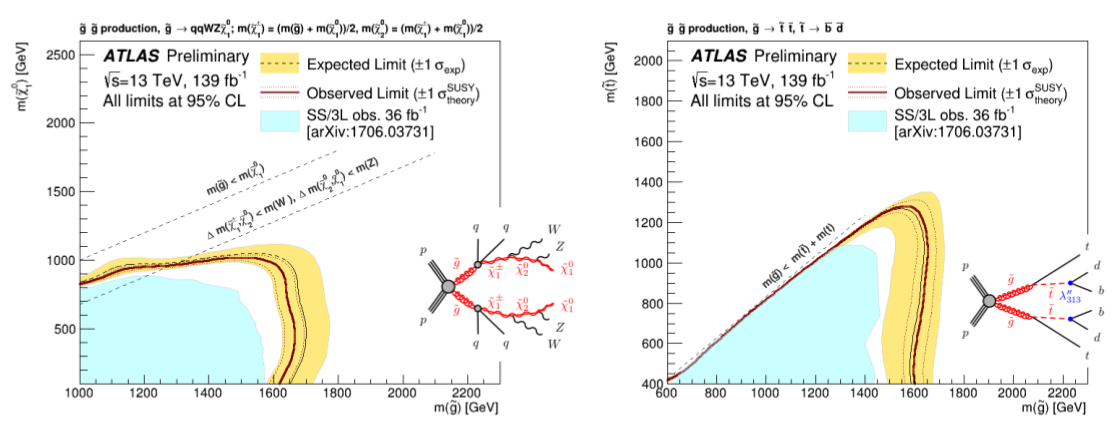}
    \caption{The 95\% confidence level exclusion limits on the production of pairs of gluinos from ATLAS. The coloured bands display the one sigma ranges of the expected fluctuations around the mean expected limit. Left plot shows the situation when the gluino pair decays into quarks and a W/Z and a neutralino, while right plot corresponds to the R-parity violating situation resulting to a top and an anti-b and an anti-d quark \cite{ATLAS-CONF-2019-015}.}
    \label{fig:Results1}
\end{figure}

In the ATLAS search with same signed leptons, the limit plot for the scenario of pair produced gluinos decaying  into quarks and a W/Z and a neutralino is shown in Figure \ref{fig:Results1} left, while the right part of the plot corresponds to the R-parity violating situation resulting to a top and an anti-b and an anti-d quark.

\begin{figure}[h!]
    \centering
    \includegraphics[width=\textwidth]{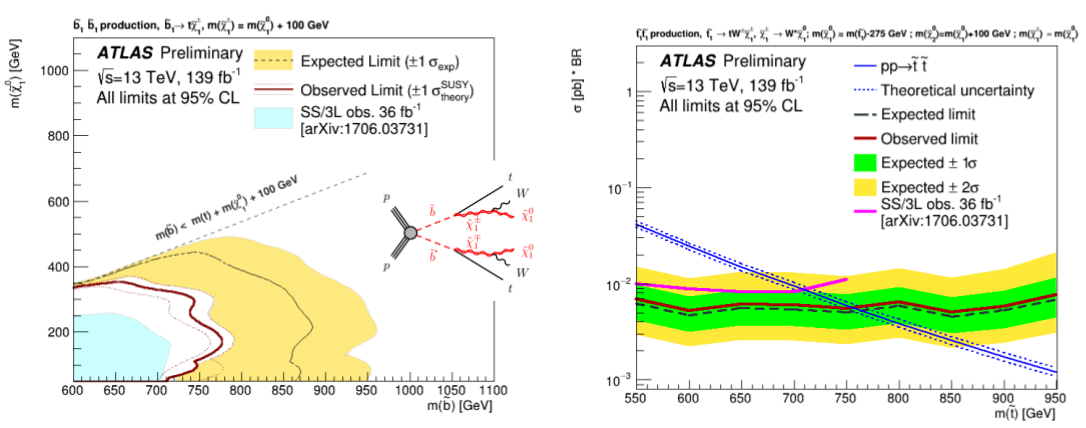}
    \caption{Exclusion limits on the production of third-generation squarks from ATLAS. The left plot shows the pair-production of bottom squarks in an R-parity conserving scenario, when the squarks decay via an intermediate chargino into a top quark, a W boson and the LSP. The right plot shows the pair-production of top squarks decaying into a top quark, a chargino and a W boson. The chargino decays to a neutralino and a W boson \cite{ATLAS-CONF-2019-015}.}
    \label{fig:ExcSquark1and2}
\end{figure}

Exclusion limits on the masses of third-generation squarks are shown in Figure \ref{fig:ExcSquark1and2}. The left plot shows the pair-production of bottom squarks in an R-parity conserving scenario, when the squarks decay via an intermediate chargino into a top quark, a W boson and the LSP. The mass of the chargino is assumed to be 100 GeV higher than the mass of the neutralino. The right plot shows the pair-production of top squarks decaying into a top quark, a chargino and a W boson. The chargino decays to a neutralino and a W boson. The masses are chosen such that the W is the dominant decay, so the mass of the top squark is 275 GeV higher than the mass of the neutralino.

Lower limits on particle masses are derived at 95\% confidence level for these models, reaching up to 1.6 TeV for gluinos and 750 GeV for bottom and top squarks. In both cases the previous results are shown as well.

\newpage

\begin{figure}[h!]
\centering
    \includegraphics[width=\textwidth]{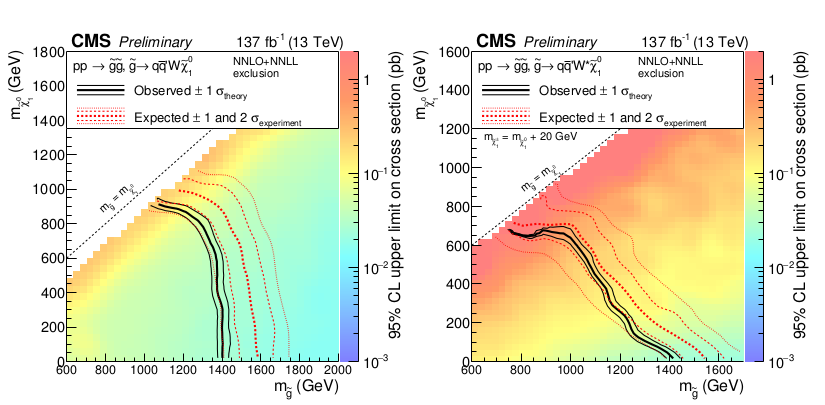}
    \caption{Exclusion regions at 95\% CL in the plane of neutralino versus gluino mass for the model when the final state has 4 quarks and 2 W bosons and 2 LSP. The left plot assumes that the mass of the chargino is the mean of the mass of the gluino and the neutralino, while the right plot assumes that the chargino is 20 GeV heavier then the neutralino \cite{CMS-SUS-19-008}.}
    \label{fig:Chargino1}
    \centering
    \includegraphics[width=\textwidth]{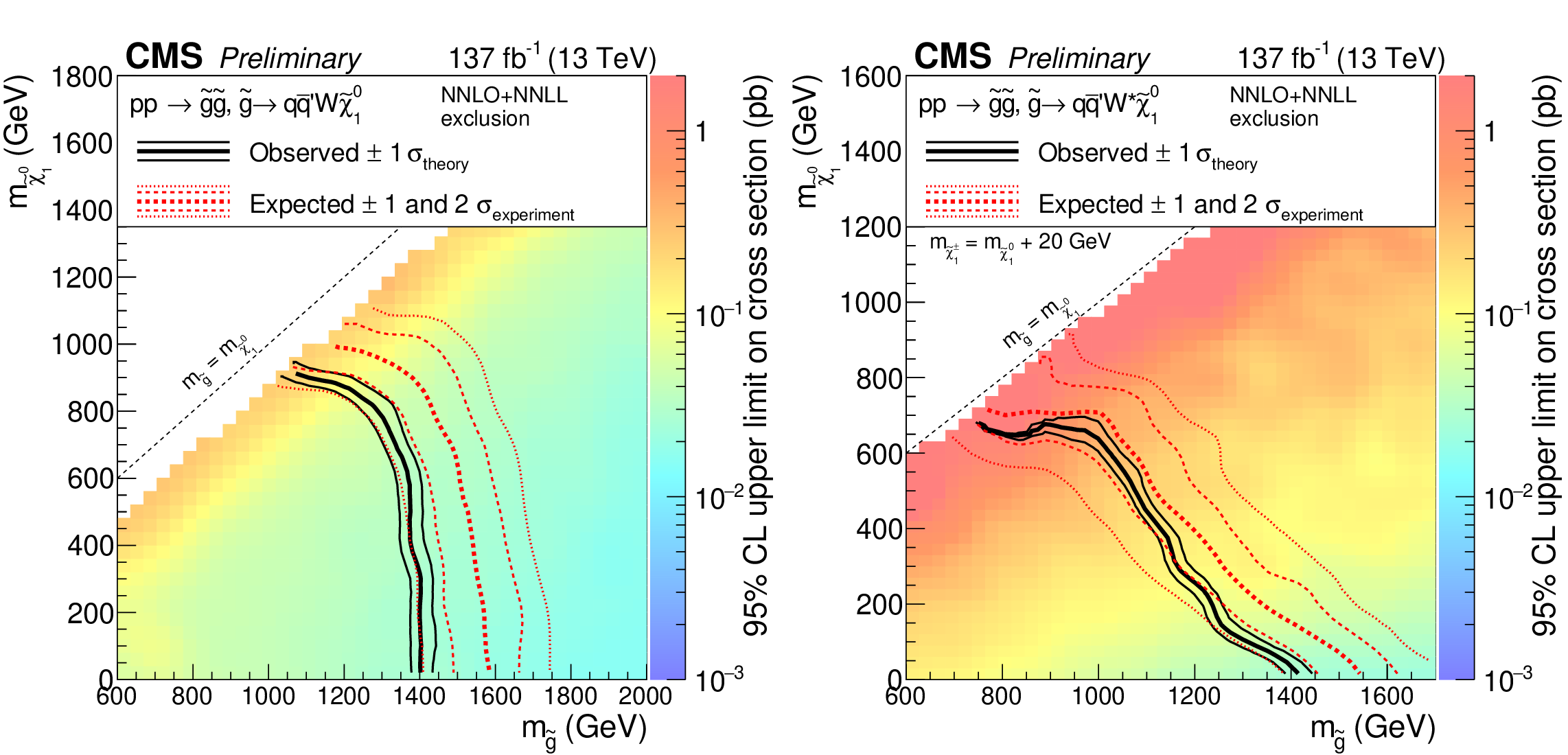}
    \caption{Limits on RPV gluino pair production with each gluino decaying into four quarks and one lepton (T1qqqqL, left) and each gluino decaying into a top, bottom, and strange quark (T1tbs, right) \cite{CMS-SUS-19-008}.}
    \label{fig:RPV1}
\end{figure}

From the CMS search with the same signature, limits on the gluino mass are shown in Figure \ref{fig:Chargino1}. The final state has four quarks and two W bosons and two LSP in that model. The left plot assumes that the mass of the chargino is the mean of the mass of the gluino and the neutralino, while the right plot assumes that the chargino is 20 GeV heavier then the neutralino.

RPV gluino pair production with each gluino decaying into four quarks and one lepton are shown on the left of Figure \ref{fig:RPV1} and each gluino decaying into a top, bottom, and strange quark is shown on the right.

Lower mass limits for the gluino are as high as 1.4 TeV for R-parity conserving scenarios, while 2.1 TeV for R-parity violating case.

\begin{figure}[h!]
    \centering
    \includegraphics[width=\textwidth]{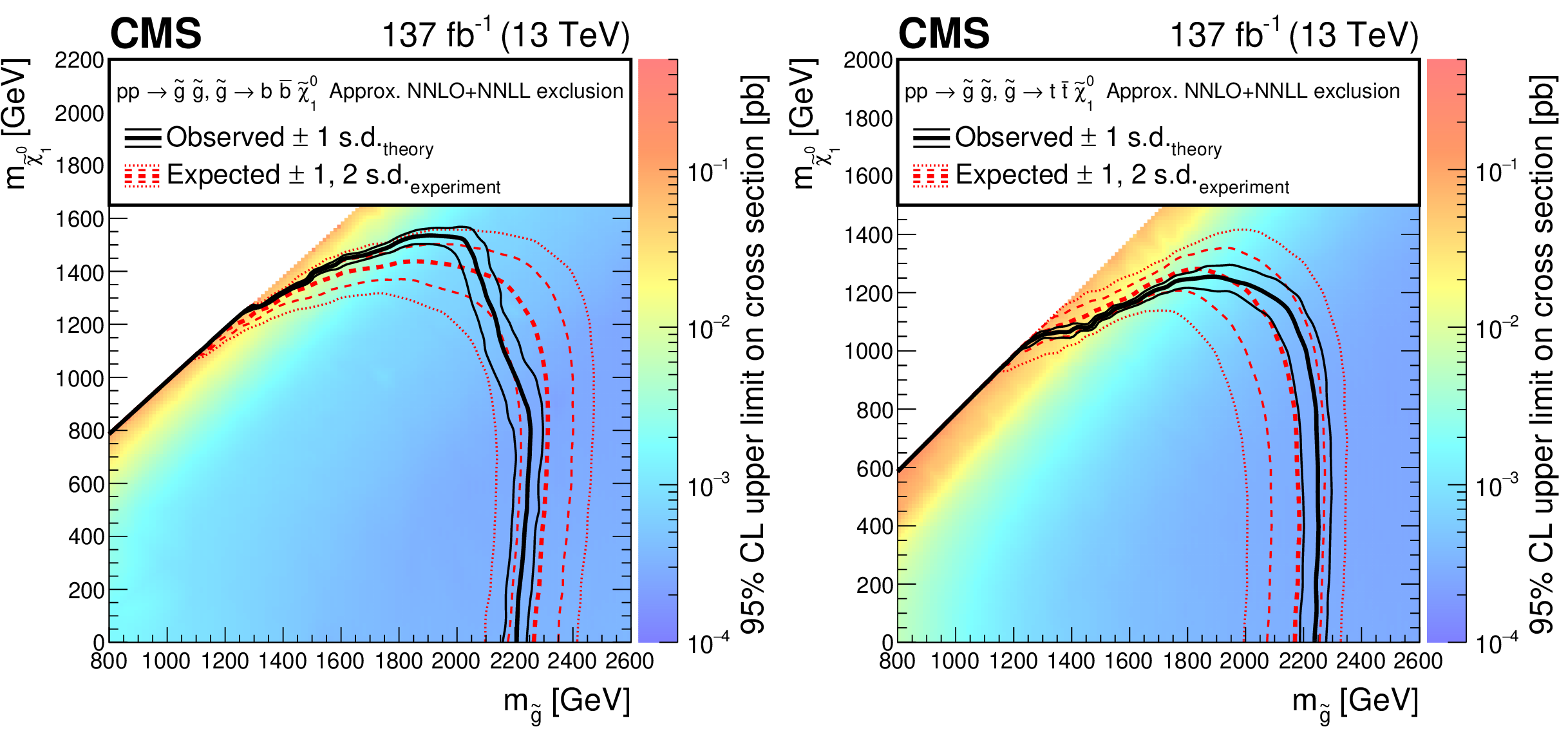}
    \caption{Exclusion limits at 95\% CL for gluino-mediated bottom squark production (left) and gluino-mediated top squark production (right) \cite{CMS-SUS-19-005}.}
    \label{fig:Excl5}
\end{figure}

\newpage
In the CMS search \cite{CMS-SUS-19-005}, in Figure \ref{fig:Excl5} exclusion limits at 95\% CL for gluino-mediated bottom squark production (left) and gluino-mediated top squark production (right) are shown. From this plot we can extract that the best exclusion limit for the masses of the gluino go up to 2250 GeV.

\begin{figure}[h!]
    \centering
    \includegraphics[width=\textwidth]{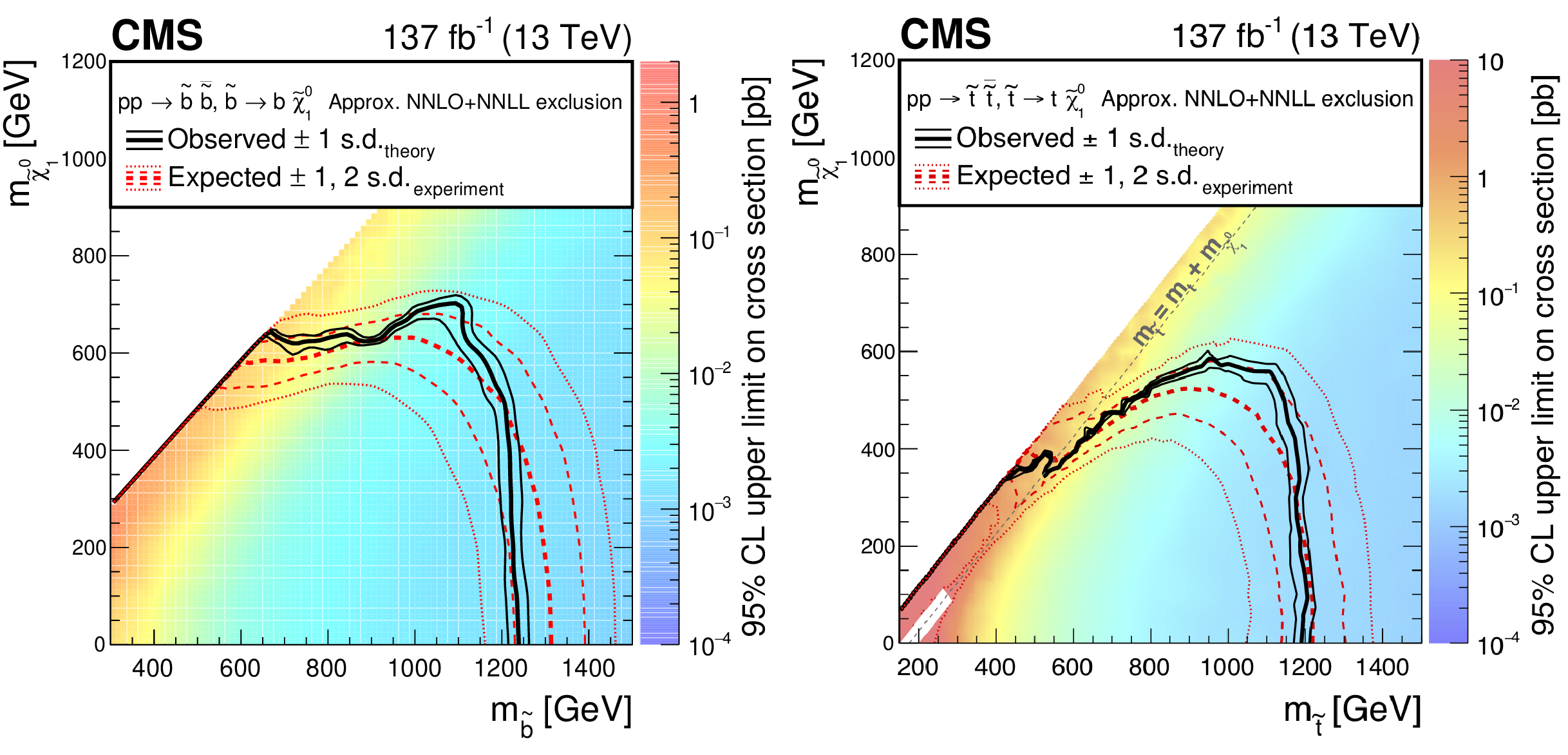}
    \caption{Exclusion limits on the production of third-generation squarks. Left plot is for the bottom squark and the right is for the top squark.  The white diagonal band in the top squark pair production exclusion limit corresponds to a region where the determination of the cross section upper limit is uncertain because of the finite granularity of the available MC samples \cite{CMS-SUS-19-005}.}
    \label{fig:ExcSquark3and4}
\end{figure}

Exclusion limits on the production of third-generation squarks is shown on Figure \ref{fig:ExcSquark3and4}. Left plot is for the bottom squark and the right is for the top squark. From these we can extract exclusion limits for the masses up to 1260 GeV for the bottom squarks and 1225 GeV for the top squarks. 

\begin{figure}[h!]
    \centering
    \includegraphics[width=\textwidth]{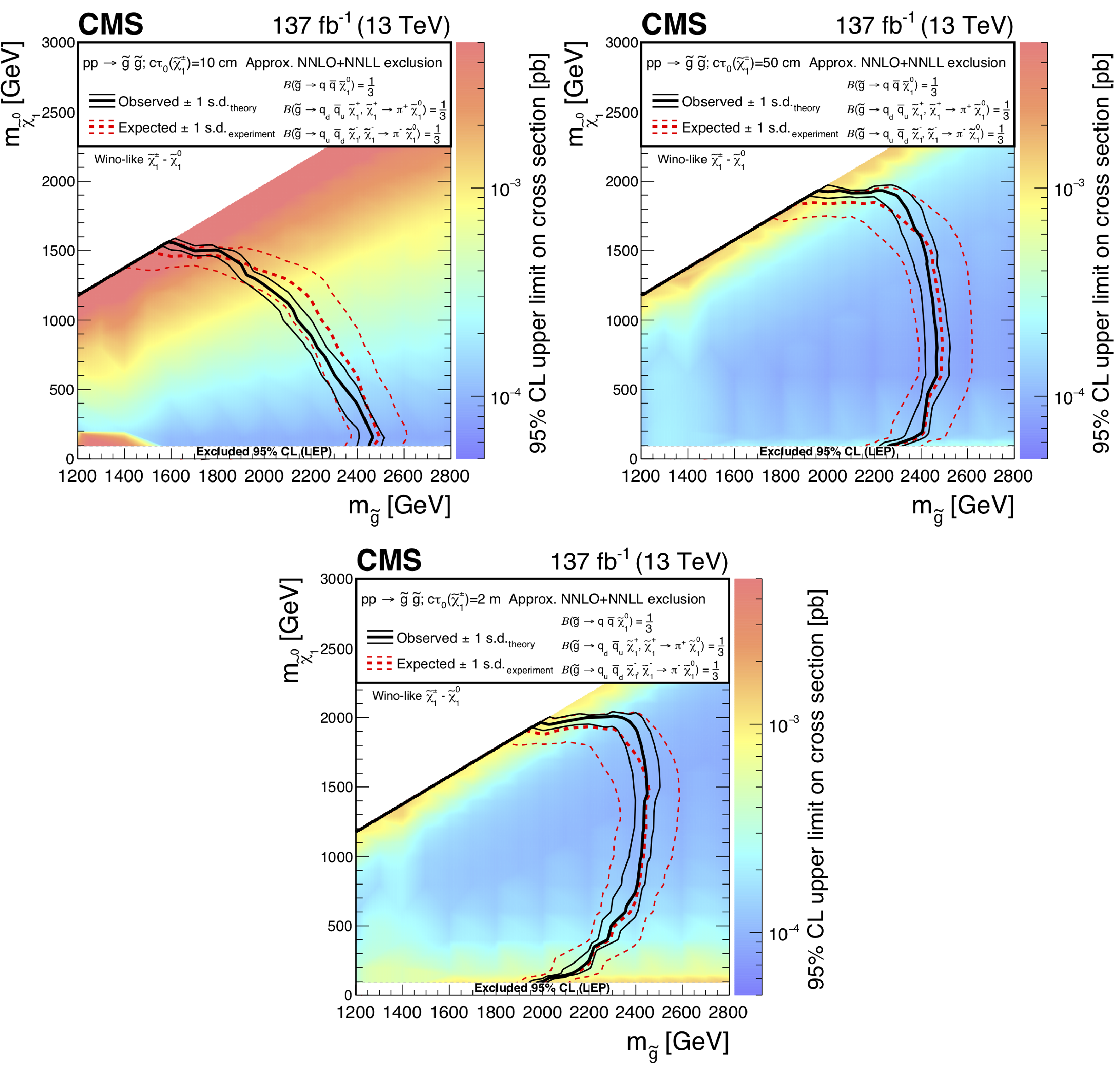}
    \caption{Exclusion limits at 95\% CL are shown for gluino-mediated light-flavor (u,d,s,c) squark production with $c\tau$ = 10 cm (above left), 50 cm (above right), and 200 cm (below) \cite{CMS-SUS-19-005}.}
    \label{fig:Excl6}
\end{figure}

If we add a disappearing track to the analysis, in Figure \ref{fig:Excl6}, exclusion limits at 95\% CL are shown for gluino-mediated light-flavor (u,d,s,c) squark production with $c\tau$ = 10 cm (above left), 50 cm (above right), and 200 cm (below). The white band in the bottom of each plot corresponds to the phase space already excluded by LEP. From these plots we can extract mass limits for the gluino and neutralino up to 2460 GeV and 2000 GeV, respectively.

\newpage
\section{Conclusions}
Three searches for strong production Supersymmetry with a total integrated luminosity of 137 fb$^{-1}$ at a center-of-mass energy of 13 TeV, corresponding to years 2015-2018 were presented. There has been no significant excess observed. The data gave the following best 95\% exclusion limits for the masses of squarks and gluinos: 2250 GeV, 1260 GeV and 1225 GeV, which are obtained from the inclusive $M_{T2}$ search for gluinos (Figures \ref{fig:Excl5} and \ref{fig:ExcSquark3and4}), bottom squarks and top squarks, respectively. The search for disappearing tracks extends the gluino mass limit to as much as 2460 GeV, and the neutralino mass limit to as much as 2000 GeV as shown on Figure \ref{fig:Excl6}. 


\end{document}